\documentclass[10pt, twocolumn]{article}
\pdfoutput=1 

\usepackage[pdftex]{graphicx}
\usepackage{subfigure}
\usepackage{threeparttable}

\usepackage{times}
\usepackage{lineno}

\usepackage{url}
\let\OLDthebibliography\thebibliography
\renewcommand\thebibliography[1]{
  \OLDthebibliography{#1}
  \setlength{\parskip}{0pt}
  \setlength{\itemsep}{0pt plus 0.3ex}
}

\title{Constructing cities, deconstructing scaling laws}

\author
{Elsa Arcaute,$^{1\ast\dagger}$ Erez Hatna,$^{2,1\dagger}$ Peter Ferguson,$^{1}$ Hyejin Youn,$^{3}$ \\Anders Johansson,$^{4,1}$ Michael Batty $^{1}$\\
\\
\normalsize{$^{1}$Centre for Advanced Spatial Analysis (CASA), University College London, UK}\\
\normalsize{$^{2}$Center for Advanced Modeling, The Johns Hopkins University, USA}\\
\normalsize{$^{3}$The Institute for New Economic Thinking at the Oxford Martin School, University of Oxford, UK; and Santa Fe Institute, USA}\\
\normalsize{$^{4}$Department of Civil Engineering, University of Bristol, UK}\\
\\
\normalsize{$^\ast$To whom correspondence should be addressed; E-mail:  e.arcaute@ucl.ac.uk.}\\
\normalsize{$^\dagger$Both authors contributed equally.} 
}

\date{}

\begin{document}

\maketitle 


\begin{abstract}

Cities can be characterised and modelled through different urban measures. Consistency within these observables is crucial in order to advance towards a science of cities. Bettencourt et al have proposed that many of these urban measures can be predicted through universal scaling laws. We develop a framework to consistently define cities, using commuting to work and population density thresholds, and construct thousands of realisations of systems of cities with different boundaries for England and Wales. These serve as a laboratory for the scaling analysis of a large set of urban indicators. The analysis shows that population size alone does not provide enough information to describe or predict the state of a city as previously proposed, indicating that the expected scaling laws are not corroborated.
We found that most urban indicators scale linearly with city size regardless of the definition of the urban boundaries. However,  when non-linear correlations are present, the exponent fluctuates considerably.

\end{abstract}

\vspace{0.4cm}

Cities are the outcome of intricate social and economic dynamics, shaped by geographical, cultural and political constraints.
There is however little understanding on how all the different features interweave and co-evolve.
Certain properties such as morphological attributes, e.g. fractality of cities \cite{Batty_FractalCities,Batty_scaling2008}, Zipf distributions of city sizes \cite{Zipf1949,Jiang_Jia2011} and population growth laws \cite{Gabaix_1999,Gabaix_Proc1999,Eeckhout2004,Rozenfeld_Batty_Makse08}, seem to transcend contextual constraints although debate remains with respect to the universality of some of these characteristics \cite{Cristelli_Batty_Pietronero2012,Giesen_etal2010,Giesen_Suedekum2011}.

\begin{figure}[h]
\centerline{\includegraphics{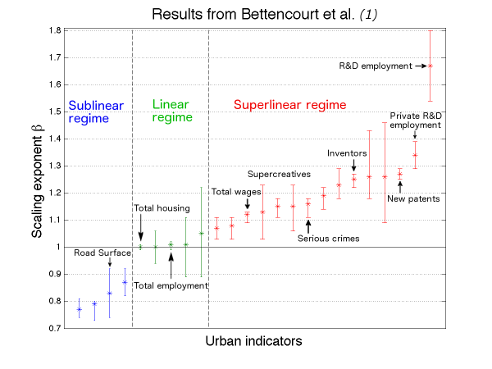}}
   \caption{Exponents with $95\%$ CI for different urban indicators found for the US, Germany and China in \cite{Bettencourt_PNAS07}. These are colour-coded according to their regime.}\label{betas_Bettencourt}
  \end{figure}

In the last decade, drawing from an analogy with Kleiber's law \cite{Kleiber,West_Science1997} which stipulates allometric scaling of the metabolic rate with respect to the mass of an animal, it has been proposed that most urban indicators can be determined in terms of the following ubiquitous scaling law \cite{Bettencourt_PNAS07,Bettencourt_etal10,Bettencourt_West2010,BettencourtScience2013}
\begin{equation}
Y(t)=Y_0(t)N(t)^\beta
\label{scalingEq}
\end{equation}
where $Y(t)$ and $N(t)$ represent the urban indicator and the population size of a city at time $t$ respectively, and $Y_0(t)$ is a time dependent normalisation constant. 
It is conjectured that the nature of the urban observable will unequivocally define one of the three universal categories to which the scaling exponent $\beta$ belongs: (i) $\beta<1$, a sublinear regime given by economies of scale associated with infrastructure and services, e.g. road surface area; (ii) $\beta\approx1$, a linear regime associated with individual human needs, e.g. housing and household electrical consumption; and (iii) $\beta>1$, a superlinear regime associated with outcomes from social interactions, e.g. number of patents and income \cite{Pumain_Paulus_Vacchiani2009}. 
Observations in the US, Germany and China \cite{Bettencourt_etal10} seem to provide empirical evidence supporting the conjectured values for the exponent in eq.~\ref{scalingEq}. These results together with their confidence intervals (CI) are pictured in Fig.~\ref{betas_Bettencourt}. These are punctual values for a single pre-determined definition of urban areas: Metropolitan Statistical Areas (MSA) in the US, and Larger Urban Zones (LUZ) in Europe. These definitions were designed  to incorporate urbanised and economic functional areas but they are not necessarily consistent with one another as  no consensus exists on how cities should be defined.

In this article we investigate the extent to which in England and Wales (E\&W)\footnote{We exclude from the analysis the other two countries of the United Kingdom in order to avoid inconsistencies between the different datasets.
The Office for National Statistics collects census data for England and Wales only, while two other different agencies collect data for Scotland and Northern Ireland separately.}, urban indicators can be estimated on the basis of size alone according to eq.~\ref{scalingEq}, regardless of constraints, such as inter-city interactions, globalisation, or simply historical dependencies. 
Instead of limiting the analysis to a single predefined definition of cities such as LUZ, we define a simple procedure that produces a system of cities by aggregating small statistical units. We chose this approach for the following reasons: (i) the LUZ  selection of cities is very small as only 21 cities in E\&W are considered while important cities such as Oxford and Reading are missing, leading to a small sample space; (ii) the procedure can be easily reproduced in other countries and it thus allows for a consistent  comparison with other urban systems,  and more importantly, (iii) this methodology provides a set of different realisations of the urbanised space, serving as a laboratory to explore the sensitivity of the urban indicators to a comprehensive set of different city and metropolitan area demarcations in E\&W, leading to a more rigorous framework to test urban hypotheses.
For the curious reader that is interested in a direct comparison with the LUZ definition, the results of the scaling analysis can be found in Fig.~S8 of the SM. The findings for LUZ do not corroborate the expected behaviour reported in \cite{Bettencourt_PNAS07}.

There are different methods to re-construct urban systems, for example through urban growth \cite{Rozenfeld_Batty_Makse08,Rozenfeld_Gabaix_Makse2011,Rybski_PhysRevE87_2013,Frasco_etal_PhysRevX2014}, or other methods using percolation and diffusion limited aggregation \cite{ Makse_Havlin_Stanley_Nature1995,Makse_Batty_Havlin_Stanley1998,Murcio_etal_Chaos2013,Fluschnik_etal_arXiv2014}. 
In this paper we apply a simple methodology that consists of two steps. The first step uses a clustering algorithm parametrised by population density. This gives rise to settlements defined through urban morphology only. For a particular range of the population density threshold, a good representation of the extent of cities can be recovered. Nevertheless, we do not limit our analysis to this range, so that we are able to analyse the robustness of the scaling exponent to the different configurations of the urban extent.

The second step consists of defining metropolitan areas based on the clusters that were obtained in the first step. This is achieved by adding areas to cities according  to a commuting threshold. The approach is similar to  the way other definitions of metropolitan areas, such MSA, are defined but instead of using a single commuting threshold (such as  the typical value of about $30\%$) we once again
define cities over the whole range of commuting thresholds.

We present the results for plausible cases of cities and metropolitan areas as well as for the entire  range of density and commuting  thresholds.

\subsection*{Data}
Most of the variables used in the analysis come from the 2001 UK census dataset, produced by the Office for National Statistics. 
The data is given at the level of wards which are the smallest geographical units in the census data across many variables. E\&W consists of 8850 wards that reflect the political geography of the country at a fine resolution and have similar populations due to the need to maintain equality of representation in political elections.

Data on household income was taken from the UK census experimental statistics for 2001/02, and it corresponds to estimates produced using a model-based process. 
Infrastructure data, such as the area of roads, paths and buildings, come from the 2001 Generalised Land Use Database. Finally, data on patents was provided by the intellectual property office at the postcode level, for the years 2000 to 2011.  Each of the tables from which the indicators were obtained is described in detail in the SM.

\subsection*{Clustering through density thresholds: cities}

\begin{figure}[ht]
\centerline{\includegraphics[width=0.5\textwidth]{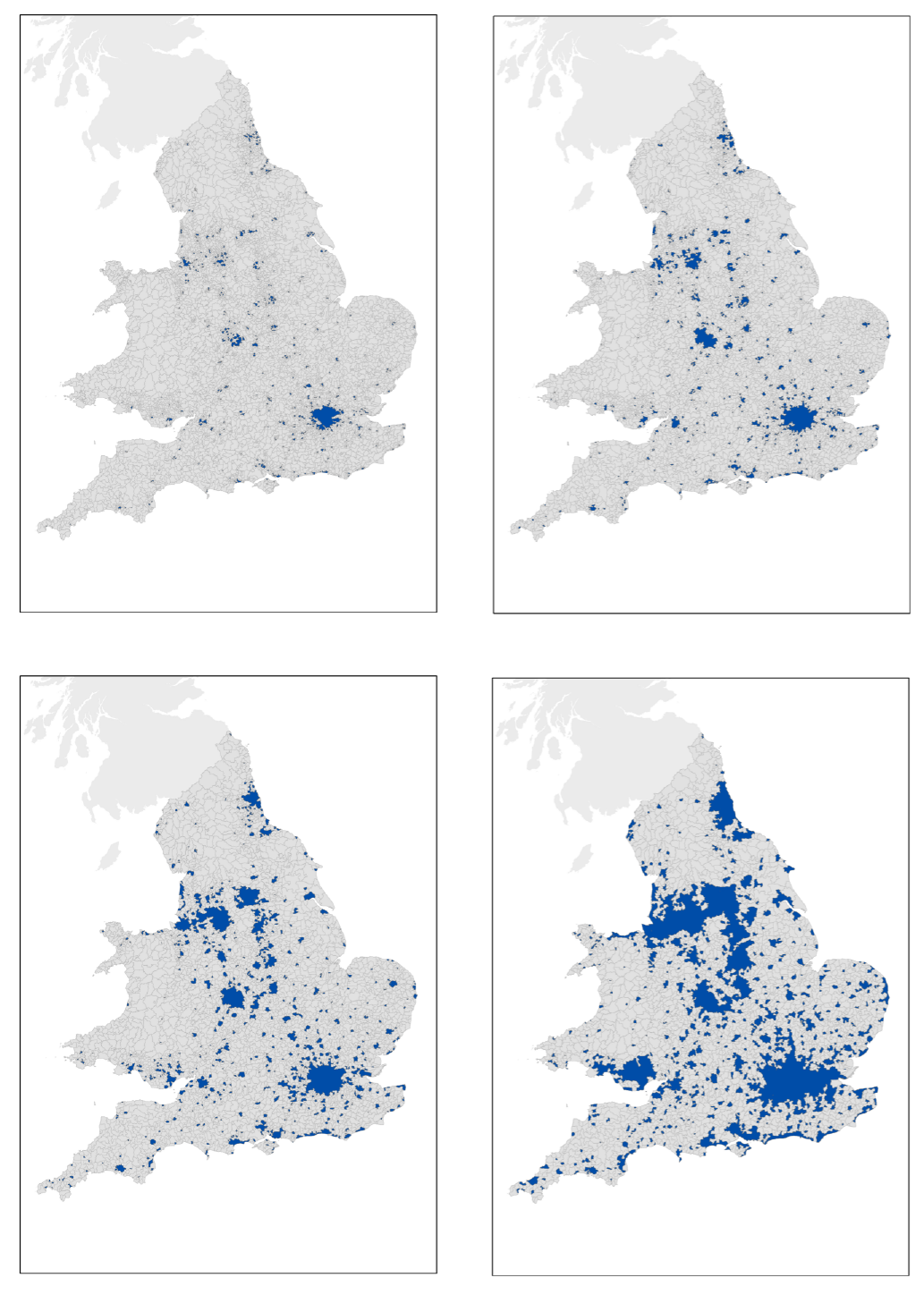}}
   \caption{Sample of configurations of cities for four different density cutoffs. From top left to bottom right: $\rho=40$prs/ha, $\rho=24$prs/ha, $\rho=10$prs/ha and $\rho=2$prs/ha.}\label{map1_40}
  \end{figure}

The algorithm described in this section gives rise to configurations of clusters representing cities and smaller settlements in terms of their morphological extent. We use population density as the main parameter, since this is an intrinsic property of urbanised spaces.
The unit of agglomeration for our algorithm is a \emph{ward} (see SM for details).
We define the parameter for population density $\rho_0$ to lie within the interval $[1;40]$ persons/hectare. For each integer threshold $\rho_0$ in the interval, we cluster all adjacent wards with density $\rho_w$ such that $\rho_w \geq \rho_0$. If a ward $k$ has a density $\rho_k < \rho_0$, but is surrounded by wards such that for each ward $w$, $\rho_w \geq \rho_0$, then the ward $k$ is also included in the cluster. The resulting city area is hence a continuous surface.
We obtain 40 different realisations of systems of cities for E\&W, varying from very large clusters containing various settlements, to clusters containing only the core of cities for the highest density values (Fig.~\ref{map1_40}).
 
\begin{figure}
\centerline{\includegraphics[width=0.5\textwidth]{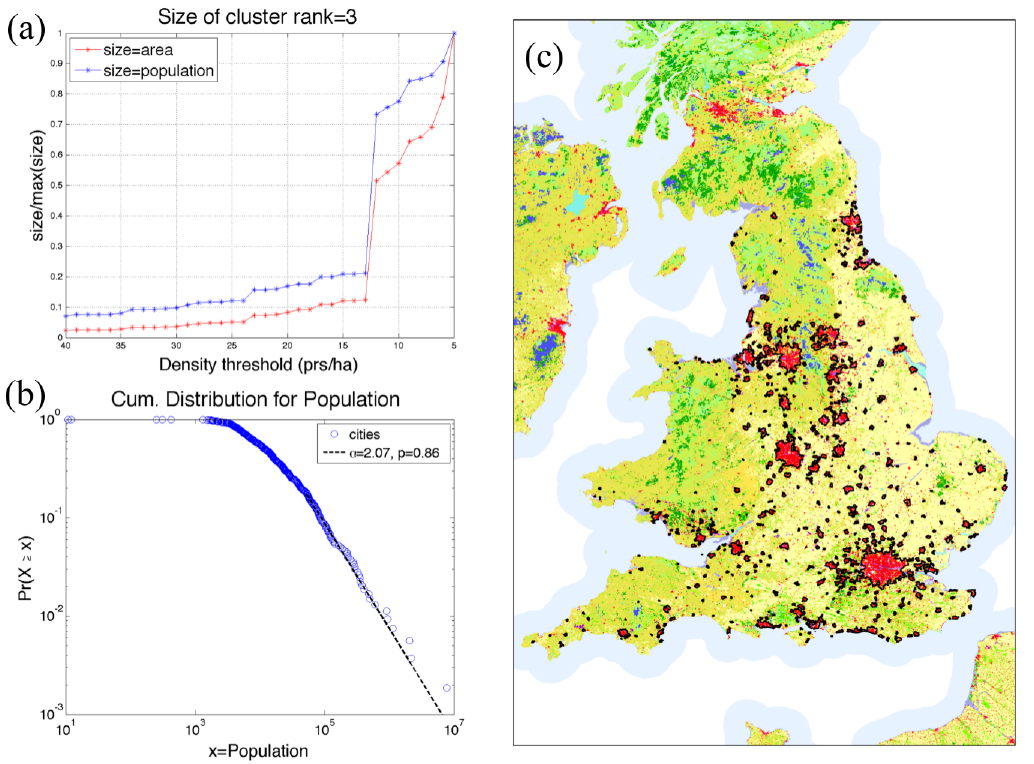}}
   \caption{System of cities defined at a density cutoff  of $\rho_c=14$prs/ha. (a) Transition of cluster size; (b) Zipf distribution of city size; (c) The Corine land cover map of E\&W: red corresponds to the built area and the black contours to the clusters defined for $\rho_c=14$prs/ha.}\label{cluster14_all}
  \end{figure}

For a range of densities, the algorithm produces realisations that are in very good agreement with the identified urbanised areas. One of many possible good realisations can be determined by looking at transitions in the cluster sizes resulting from the change in density from high to low values. 
The largest cluster exhibiting a sharp transition is the third biggest one (rank 3, see Fig.~\ref{cluster14_all}(a)), and the jump corresponds to the merging of Liverpool and Manchester. Given that these two cities are very close, we select the density threshold $\rho_c=14$prs/ha, which is near the transition and before the joining takes place.  
It is important to note that this choice is not unique, and the properties and results that we will show below hold for a range of choices of $\rho$.
The system of cities defined at $\rho_c$ has a Zipf distribution of city sizes\footnote{The exponent was computed using the method for fitting a power-law distribution proposed in \cite{Clauset_etalPL09}.}, Fig.~\ref{cluster14_all}(b), and the boundaries, displayed as black contours in Fig.~\ref{cluster14_all}(c), show an excellent overlap with the built areas (red clusters in the map) derived from remote sensing  \cite{Corine_ref}. Cities specified at the density of $\rho_c=14$prs/ha are therefore a good proxy for a definition of cities vis \`a vis of their morphology, i.e. the urbanised space.

\subsection*{Clustering through commuting thresholds: metropolitan areas}

Metropolitan areas correspond to urban agglomerations linked together through socio-economic functionalities. 
We  construct such areas by considering the density based cities as destinations of commuter flows. For each city, we add the areas that are the origins of its commuter flows.

In order to include small settlements as origins rather than destinations of commuting flows, we impose a minimum population size on the initial clusters, such that only the larger settlements are considered commuting hubs. The data on commuter flows at the ward level is taken from the 2001 census of the UK \cite{ONS_commuting2001}.

In detail, this second algorithm works as follows.
For each density realisation $\rho_0 \in \{1,2, ...,40\}$ prs/ha, we impose a minimum population size cutoff $N\geq N_0$ for each of the clusters, where $N_0 \in \{0,10^4,5\tiny{\times}10^4,10^5,1.5\tiny{\times}10^5\}$ individuals\footnote{$N_0=0$ individuals corresponds to the case where no cutoff is imposed, and all the settlements are taken into account.}.  
We remove smaller clusters to allow their constituting wards to be part of larger clusters, as is the case of satellite settlements around London.
For every given ward, we compute the percentage of individuals commuting to each of the clusters out of the total number of commuters from the ward. The ward is added to the cluster that receives the largest flow if the flow is above a threshold $\tau_0$\footnote{If two or more clusters have the same largest flow, the ward is assigned to one of them at random.}.
We investigate all the different realisations for $\tau_0\in\{0, 5, ..., 100\}\%$ individuals commuting from a ward to a cluster. 
The extreme value of $\tau_0=100\%$ reproduces the original system without commuting as the percentage of commuters from a given ward cannot exceed 100\%. The other extreme value of $\tau_0=0\%$ in which a ward is added to a cluster if a single individual commutes to it, leads to an almost full coverage of E\&W, where nearly every ward belongs to a cluster.

\begin{figure*}[ht]
\centerline{\includegraphics[width=0.9\textwidth]{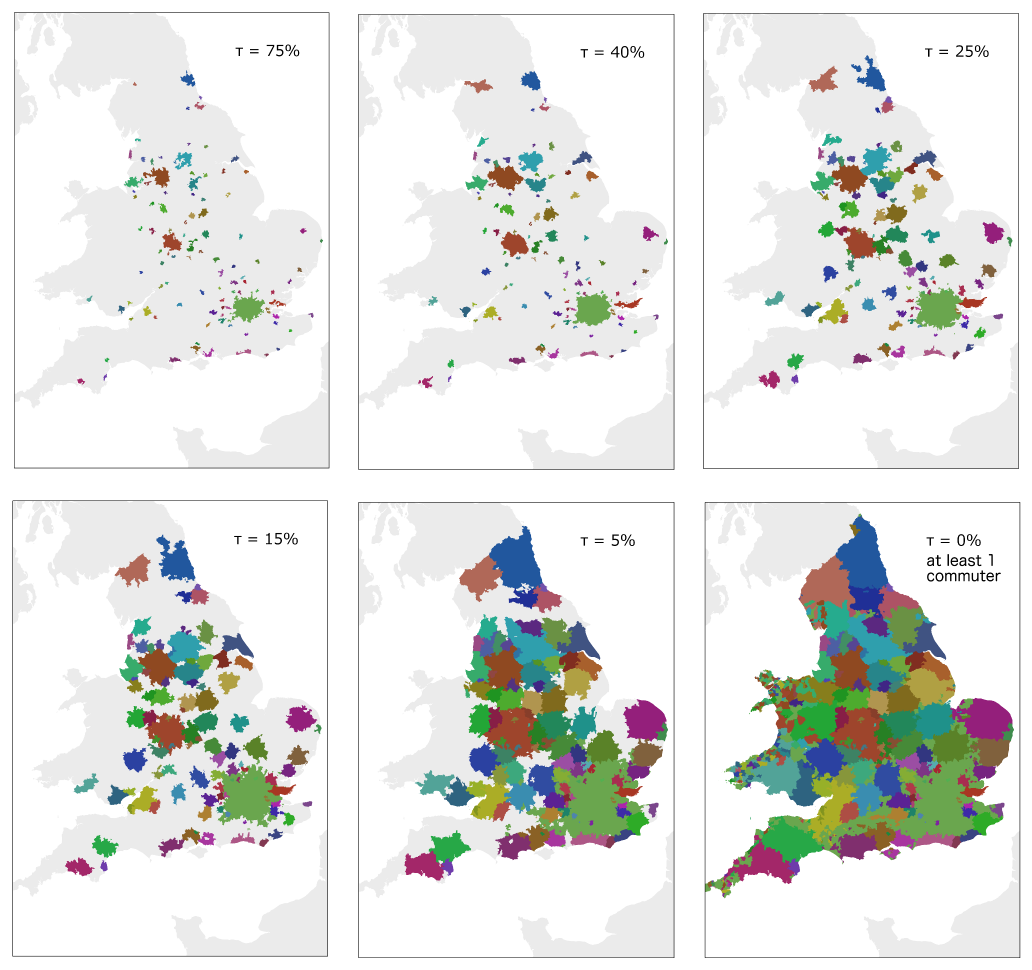}}
   \caption{Realisations of metropolitan areas at a fixed density cutoff of $\rho_c=14$prs/ha and a minimum population size of $5\tiny{\times}10^4$ individuals for a selection of several commuting flow thresholds $\tau$.}\label{comm_maps14}
  \end{figure*}

Specific realisations for the density cutoff of $\rho_c=14$prs/ha, a minimum population size of $N=5\tiny{\times}10^4$ individuals and different  flow  thresholds $\tau$, are pictured in Fig.~\ref{comm_maps14}.    
Noticeable changes in the configuration of the clusters are observed below the threshold of $50\%$, indicating that rarely the majority of individuals in a ward will commute to a single cluster. 
As a result the realisation for a flow of $75\%$ is almost identical to the realisation pre-commuting clustering.
This method gives rise to more than $2\tiny{\times}10^4$ realisations of systems of cities, that serve as a laboratory to assess the behaviour of the scaling exponent in eq.~\ref{scalingEq}.

\section*{Results for cities and metropolitan areas}

We focus in this section on the scaling analysis for cities and metropolitan areas in order to make our results comparable with other studies.
We already demonstrated that clusters defined at $\rho_c=14$prs/ha (Fig.\ref{cluster14_all}) provide a good proxy for cities, and hence we use this definition in the analysis. 
Metropolitan areas are commonly understood as cities that include the regions from which at least 30\% of the population commute to work. 
We therefore construct the metropolitan areas through the second clustering method for $\rho_c=14$prs/ha and $\tau_0=30\%$.

\begin{figure}[h]
\centering{
 	 \includegraphics{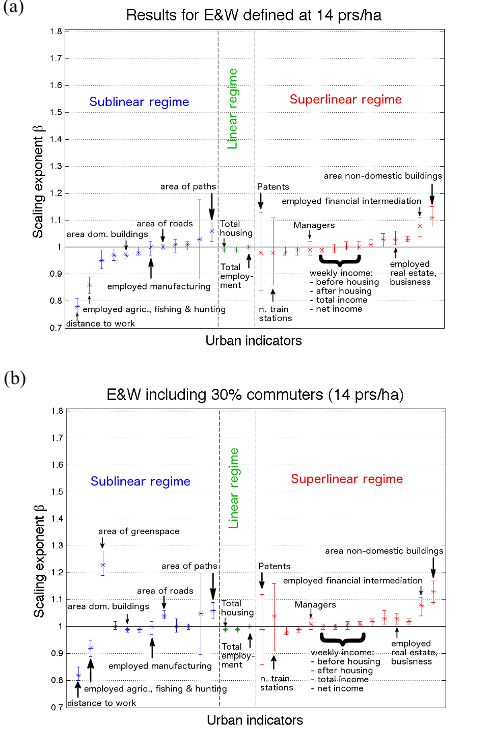}}
   \caption{Scaling exponents with $95\%$ CI for different urban indicators for cities defined a at cutoff of $\rho_c=14$prs/ha in England and Wales without commuters (a) and with 30\% commuters (b).}\label{betas_UK}
  \end{figure}

The results of the analysis are summarised in Fig.~\ref{betas_UK}(a) for cities, and in Fig.~\ref{betas_UK}(b) for metropolitan areas. The details of the variables plotted in the figures are provided in table~S1.

We observe that for most measures, any deviations of the exponent $\beta$ from linearity are extremely mild, and sometimes into the wrong regime, not corroborating the expected scaling laws. A clear illustration of this problem is given, for the sublinear regime, by the area of roads and by the area of paths; and for the superlinear regime, by the number of patents and by some employment categories, such as that for Managers. 
In the next section we will look in detail at patents, since these provide a clear example of the main issues that arise when trying to derive generic rules for urban indicators.

\subsection*{Patents}

The number of patents produced is generally considered a proxy of the city's level of innovation.  
Nevertheless, there are many cities that do not have a single patent recorded over ten years. 
Some of these cities have more than $1.8\tiny{\times}10^5$ people, while many other small ones of less than $2\tiny{\times}10^3$ inhabitants have patents registered. 

In order to investigate the resilience and urban significance of the scaling exponent for this variable, we consider two scenarios. The first scenario corresponds to the urban system containing only cities larger than $10^4$ people; and the second scenario only considers cities larger than $5\tiny{\times}10^4$ individuals. These two different population cutoffs are applied to the cities and metropolitan areas defined above.
In the literature, it is often the case that either of these two population cutoffs are employed to distinguish between a small settlement and a truly urbanised space.

\begin{figure*}[ht]
\centerline{\includegraphics[width=0.9\textwidth]{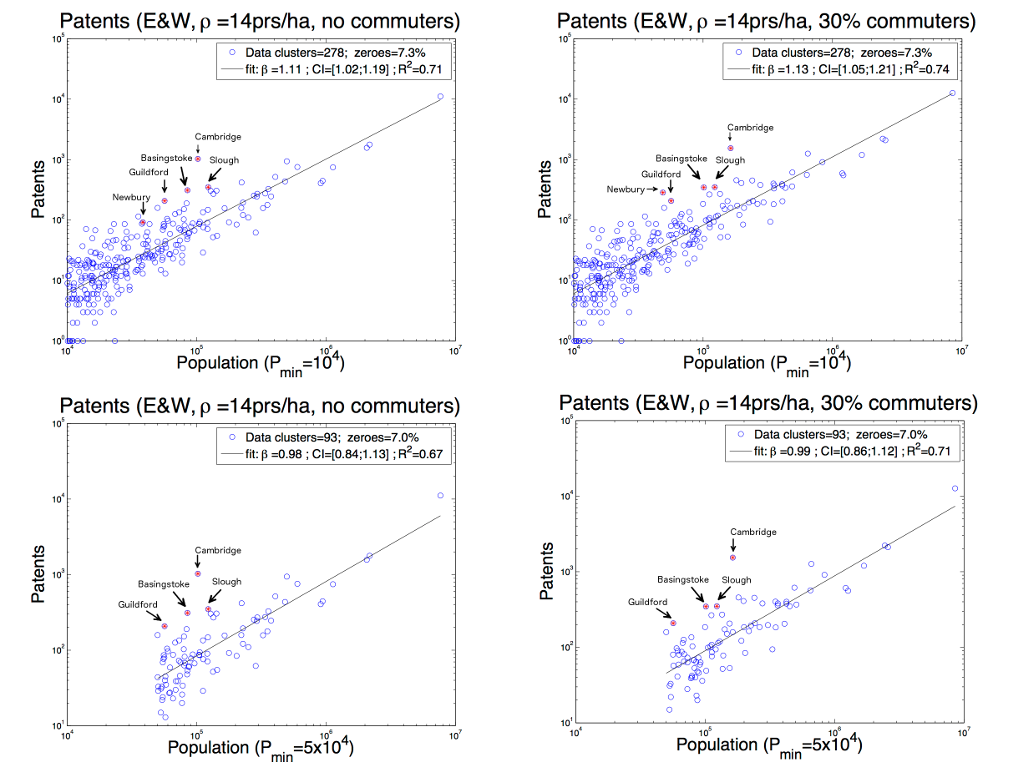}}
   \caption{Scatter plots of patents for 2 different population size cutoffs. The top plots have a minimum population size of $10^4$ individuals, while the bottom ones have a cutoff of $5\tiny{\times}10^4$ individuals.}\label{patents_clust14}
  \end{figure*}

Scatter plots of patents and city size are shown in Fig.~\ref{patents_clust14} for the two different cases.
The plots show strikingly different results for the two population cutoffs. 
For the cutoff of $10^4$ individuals, the exponent lies within the superlinear regime (at a confidence level of 95\%), while for the cutoff of $5\tiny{\times}10^4$individuals, linearity cannot be rejected.
The absence of robustness for the scaling exponent for these two cases suggests that there is a lack of self-similarity for the full range of scales examined.
This brings into question whether a minimum population size for settlements should or not be considered. Such a behaviour is often observed in systems that present power laws only for the tail of their distribution. 
Nevertheless, in this case this variable has zero values for many of the clusters, leading to a substantial amount of zero counts, including cities as large as of the order of $10^5$ individuals. 
These are given in the form of percentages in the plot.

The sensitivity of the scaling exponent to the population size cutoff indicates on the one hand, that the value of the exponent can bear no real significance on the behaviour of the system. 
And on the other, this urban indicator is unable to present a quantifiable measure over 10 years for some large cities.
This suggests that such a measure might be inadequate to properly quantify innovation.

In addition the plots indicate that the most productive cities relative to size are not the biggest ones but places that are highly rooted in education, such as Cambridge and Guildford or places corresponding to technological and business hubs. The latter are strategically located in the M4 corridor: eg Newbury (headquarters of vodafone) and Slough (the largest industrial and business estate and headquarters of Telefonica 02), or are equally well connected into other strategic transport links within the Greater South East \cite{Reades_Smith2013}, such as Guildford (in addition to the university it is also the headquarters of Philips) next to the M25 and Basingstoke (headquarters of many telecommunication companies) next to the M3.
In this case it is clear that in order to assess performance, one needs to go beyond size and consider path dependencies.

\section*{Sensitivity analysis}

In this section we look at the sensitivity analysis of the scaling exponent $\beta$, to the different boundaries of cities and metropolitan areas.
We make use of heatmaps to illustrate the value of $\beta$, where the horizontal axis represents the parameter for the density threshold, and the vertical axis the parameter for the percentage of commuters in the clustering algorithms.

\begin{figure}[h]
\centering{\includegraphics{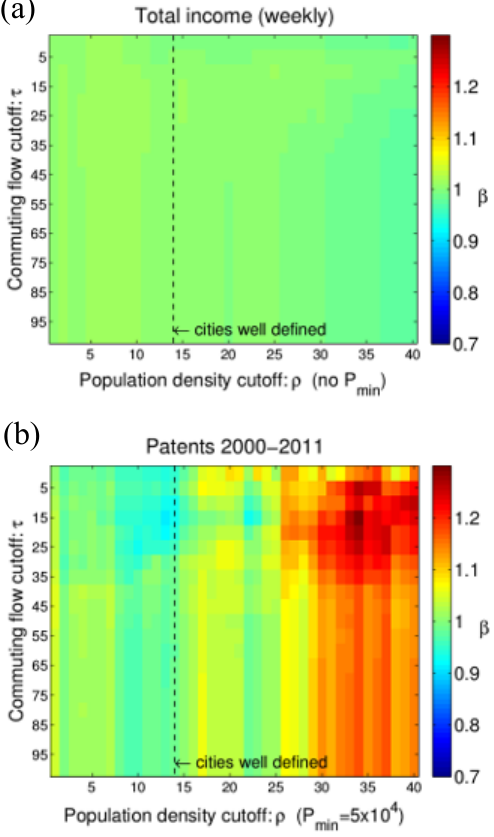}}
   \caption{(a) Heatmap for total income (no minimum population size imposed); (b) Heatmap for total number of patents for cities bigger than $5\tiny{\times}10^4$ people.}\label{heatmaps_UK}
  \end{figure}

The heatmap  in Fig.~\ref{heatmaps_UK}(a) clearly shows that for total income, population size does not convey any information on agglomeration effects, showing homogeneity throughout the map for all the different city demarcations. 
The same results were found for many other variables where superlinear exponents were expected, such as employment categories reflecting economic activity or requiring highly skilled individuals (see SM for more examples).
On the other hand, the heatmap in Fig.~\ref{heatmaps_UK}(b) shows that for variables that present non-linear dependencies, such as patents, the scaling exponent is sensitive to boundary definitions.

\section*{Discussion}

We showed that for all the different definitions of cities and metropolitan areas devised with our methodology, population size does not fully grasp the economic intricacies that constitute a system of cities in E\&W.  
Looking at possible causes of discrepancy between our results and those previously found, it is evident that London plays a special role within the urban system of the UK. 
Its strong role as an information and economic hub suggests that the urban system is highly integrated and that it is difficult to partition the system into individual cities that capture these social interaction effects.
On the other hand, if these economic functional areas are integrated following our methodology, we observe that for most urban indicators, London over-performs with respect to all other places in E\&W. 
Its position as a primate city \cite{Jefferson1939}, but most importantly, as a world city in a relatively small country, could be affecting the entire urban system. 
The performance of cities such as London should possibly be evaluated relative to other global hubs operating within a larger scaled network of interactions. 
Following Sornette's idea on the emergence of `big things' \cite{Sornette2009,Yukalov_Sornette2012,Pisarenko_Sornette2012}, a different perspective on the description of cities could be adopted, in which these global hubs are evaluated separately to their domestic counterparts. Sornette refers to the former as \emph{dragon-kings}. A statistical test showing that London can be classified as such can be found in the SM. A two-system theory of cities might then emerge: a regime for cities driving international dynamics, the dragon-kings, and a regime for the remaining cities composing a country.

In addition to economic hubs, one also encounters knowledge hubs, which also present dragon-king like qualities and which are not necessarily correlated with size. These hubs are the outcome of path-dependencies that give rise to emergent properties that are not present in all cities as is the case of patents. 
This is most dramatically demonstrated by the dominance of patent production in Cambridge, UK. 

There are many difficulties in measuring the performance of a city through scaling laws. As discussed, there are problems in defining innovation in terms of patent counts, and this is not a unique case, other variables, such as $CO_2$ emissions, present conflicting results. Some studies have found a sublinear relationship while others have found a superlinear relationship between $CO_2$ emissions and city size \cite{Fragkias_etal_PlosOne2013,Rybski_CO2_arXiv2013,Oliveira_SciRep2014,Louf_Barthelemy_SciRep2014}. Such differences might stem from the nature of the measurement itself, whether the study refers to total or only transport emissions, and/or from qualitative differences between systems such as a country's level of development.

All this indicates that a theory of cities cannot rest simply on a relationship like eq.~\ref{scalingEq}, since relevant patterns pertaining to social behaviour, such as the well-known Pareto distribution of wealth, cannot be grasped if only aggregated values are considered.
A theory of cities needs therefore to reproduce the main relevant emergent behaviours which are encoded in the diversity and heterogeneities of cities. 
It is only through this perspective that city planning and policy making can be effective.

\section*{Acknowledgements}
EA, EH, PF, AJ and MB acknowledge the support of ERC Grant 249393-ERC-2009-AdG. 
EH acknowledges support from J. M. Epstein's NIH Director's Pioneer Award, number DP1OD003874 from the National Institute of Health.
Useful discussions with Luis Bettencourt, and Geoffrey West of the Santa Fe Institute, and Jos\'e Lobo of Arizona State University helped clarify many issues. 
HY acknowledges the support of grants from the Rockefeller Foundation and the James McDonnell Foundation (no. 220020195).


\newpage

\fontsize{9pt}{12pt\selectfont}

\renewcommand{\thefigure}{S\arabic{figure}}
\renewcommand{\thetable}{S\arabic{table}}
\setcounter{figure}{0}

\topmargin 0.0cm
\oddsidemargin 0.2cm
\textwidth 16cm 
\textheight 21cm
\footskip 1.0cm

\newcommand{\beq}{\begin{equation}}
\newcommand{\eeq}{\end{equation}}

\onecolumn 
  \centerline{\huge Supplementary Materials}
  \vspace{3ex}

\setlength\parindent{0pt}

{\bf Abbreviations:} E\&W, England and Wales; CAS, Census Area Statistics; ST, Standard Table; SIC, Standard Industrial Classification
\section*{Unit of Geography}
The underlying spatial unit for all city cluster aggregations is the Census Area Statistics (CAS) ward definition produced by the UK Office for National Statistics. Ward boundaries reflect the political geography of the UK at a fine resolution and due to the need to maintain equality of representation in political elections, have similar populations. CAS ward boundaries in particular have been the standard format for the release of ward level census information since 2003. They reflect electoral ward boundaries promulgated as at 31/12/2002 and contain 8850 separate wards for England and Wales \cite{ONS1,ONS2}.

Census data used in the study was only available for England and Wales as the process for collating data in Scotland and the definition of geographic boundaries meant that equivalent datasets could not be produced. More information on census geography for Scotland for the 2001 census can be found at \cite{Census_Scotland}.

\section*{Datasets}
This section outlines the details of the datasets used in our research. The source for each of the variables together with their correspondent code is described. When referring to census data, this corresponds to the 2001 UK census produced by the Office for National Statistics.

The original data and associated metadata for tables UV02, UV53, KS15, UV34, KS12 and household income can be found under the topics section of the UK neighbourhood statistics website \cite{ONS1}.
\subsection*{Population  (Table UV02)}
The data on population was taken from the UK census table UV02. 
The data is given at the CAS ward level, and provides separate statistics for total population, ward area and a result in population density figure. The total population figure was used for all regressions with socio-demographic variables used in the study. 

\subsection*{Housing Stock (Table UV53)}
Data on household dwelling numbers comes from the census table UV53. 
The table provides information on the number of households, occupied or unoccupied, within each ward. Unoccupied household spaces are split into second residences/holiday accommodation, and vacant household spaces. A household space is the accommodation occupied by an individual household or, if unoccupied, available for an individual household. The population of this table is therefore all household spaces. The category used for regression was all household spaces and therefore included all spaces whether they were occupied or unoccupied.

\subsection*{Travel to Work (Table KS15)}
Data on travel to work distances was taken from the UK census table KS15. 
The table shows both the length and the means of travel to work used for the longest part, by distance, of the usual journey to work. For the purposes of this table, public transport is defined as underground, metro, light rail or tram, train and bus, minibus or coach. The distance travelled to work is the distance in kilometres of a straight line between the residence postcode and workplace postcode. The distance is not calculated for people working mainly at or from home, people with no fixed workplace, people working on an offshore installation or people working outside the UK. The population of the table is all people aged 16 to 74 in employment.

\subsection*{Industry of Employment (Table UV34)}
Data on the industry of employment of employees was taken from the UK census table UV34. 
The table shows the usual resident population aged 16 to 74 in employment by the industry they work in. The industry in which a person works is determined by the response to the 2001 census question asking for a description of the business of the person’s employer (or own business if self-employed). The responses were coded to a modified version of the UK Standard Industrial Classification of Economic Activities 1992 –UK SIC (92). 

In the 2001 census, industry of employment information was collected for usual residents. A usual resident was generally defined as someone who spent most of their time at a specific address. It includes: people who usually lived at that address but were temporarily away (on holiday, visiting friends or relatives, or temporarily in a hospital or similar establishment); people who worked away from home for part of the time; students, if it was their term-time address; a baby born before 30 April 2001 even if it was still in hospital; and people present on census day, even if temporarily, who had no other usual address. However, it did not include anyone present on census day who had another usual address or anyone who had been living or intended to live in a special establishment, such as a residential home, nursing home or hospital, for six months or more.

The industry of employment categories used in the study were the following: agriculture, hunting and forestry; manufacturing; construction; hotels and restaurants; financial intermediation; real estate, renting and business activities; public administration, defence and social security; education.

\subsection*{Occupational Groups (Table KS12a)}
Data on occupational groups was taken from the UK census table KS12.
The information on this table comes from responses to questions asking for the full title of the main job and a description of the job. The population includes any person aged 16 to 74 who carried out paid work in the week before the census, whether self- employed or an employee, is described as employed or in employment. 'Paid work' includes casual or temporary work, even if only for one hour; being on a government- sponsored training scheme; being away from a job/business ill, on maternity leave, on holiday or temporarily laid off; or doing paid or unpaid work for their own or family business. 

The following occupational groups were used in the regression analysis: managers and senior officials; professional occupations; associate professional and technical operations; skilled trades occupations; administrative and secretarial occupations; personal service occupations; sales and customer service occupations; process; plant and machine operatives; elementary occupations (examples of elementary occupations include farm workers, labourers, kitchen assistants and bar staff).

\subsection*{Household Income}
The dataset on household income was taken from the UK census experimental statistics for 2001/02 and is provided at a fine geographic resolution for the whole of England and Wales. The original data and associated metadata for household income can be found under the topics section of the UK neighbourhood statistics website \cite{ONS1}.

The income dataset correspond to estimates that were produced using a model-based process which involves finding a relationship between survey data (data available on income) and other data drawn from administrative and census data sources. A model fitting process was used to select co-variates with a consistently strong relationship to the survey data. The strength of the relationship with these covariates was used to provide estimates on income for those wards where survey data on income is not available. More information on the provenance of the income data can be found on the appropriate page of the UK neighbourhood statistics census access site. 

The survey data on income was taken from the Family Resources Survey (FRS) for the same year (2001/02)\footnote{The FRS is produced by the UK Department for Work and Pensions (DWP) to ensure a large sample sizes when collating information on household expenditure. Information on the FRS for 2001 can be found on the research section of the dWP.gov website \cite{dwp}, the methodology section (section 8) of the FRS summary report for that year at \cite{dwp_report}  and the associated technical report available through the ONS \cite{ONS2}}. The total sample size for the 2001 survey was 42,000 addresses taken from across the UK. 

In this study we used the average weekly household total income (unequivalised) estimations in Pounds Sterling. We converted the weekly average income of each ward into weekly sum of incomes  by multiplying the average by the number of households. The number of households within each ward was taken from the census Household Composition table (UV65).

\subsection*{Patents}
Patent data was provided by the UK Intellectual Property Office with postcode level reference that was subsequently aggregated to the CAS ward level. Data was provided for the years 2000 to 2011 inclusive to ensure sufficient quantity to avoid null values for individual wards. The total number of patents in the dataset that could be identified in E\&W was 66,270. The values used for regression were simply the gross number of patents registered to a particular postcode whether it be business or home address. 

More information on patent information from the UK IPO can be found at \cite{ipo}.

\subsection*{Land Use Statistics}
All the measures related to land use were taken from the Generalised Land Use Database, 2001 (GLUD). Below we describe the geographical units and classifications.
 
The GLUD figures show the areas of different land types for census Output Areas (OAs), Lower Layer Super Output Areas (LSOAs), Middle Layer Super Output Areas (MSOAs), Local Authorities (LAs), and Government Office Regions (GORs) in England as at 1st November 2001. Output level data was aggregated to the ward level for comparative analysis with population.

For the GLUD, a classification has been developed which allocates all identifiable land features on the UK Ordnance Survey MasterMap® national mapping product into nine simplified land categories and an additional 'unclassified' category. These are: (1) Domestic buildings (2) Non-domestic buildings (3) Roads (4) Paths (5) Rail (6) Gardens (domestic) (7) Greenspace (8) Water (9) Other land uses (largely hardstanding) and (10) Unclassified.

\section*{Scaling results for cities defined at \boldmath$\rho_c=14$prs/ha}

\begin{figure}[ht]
\centerline{\includegraphics[width=0.8\textwidth]{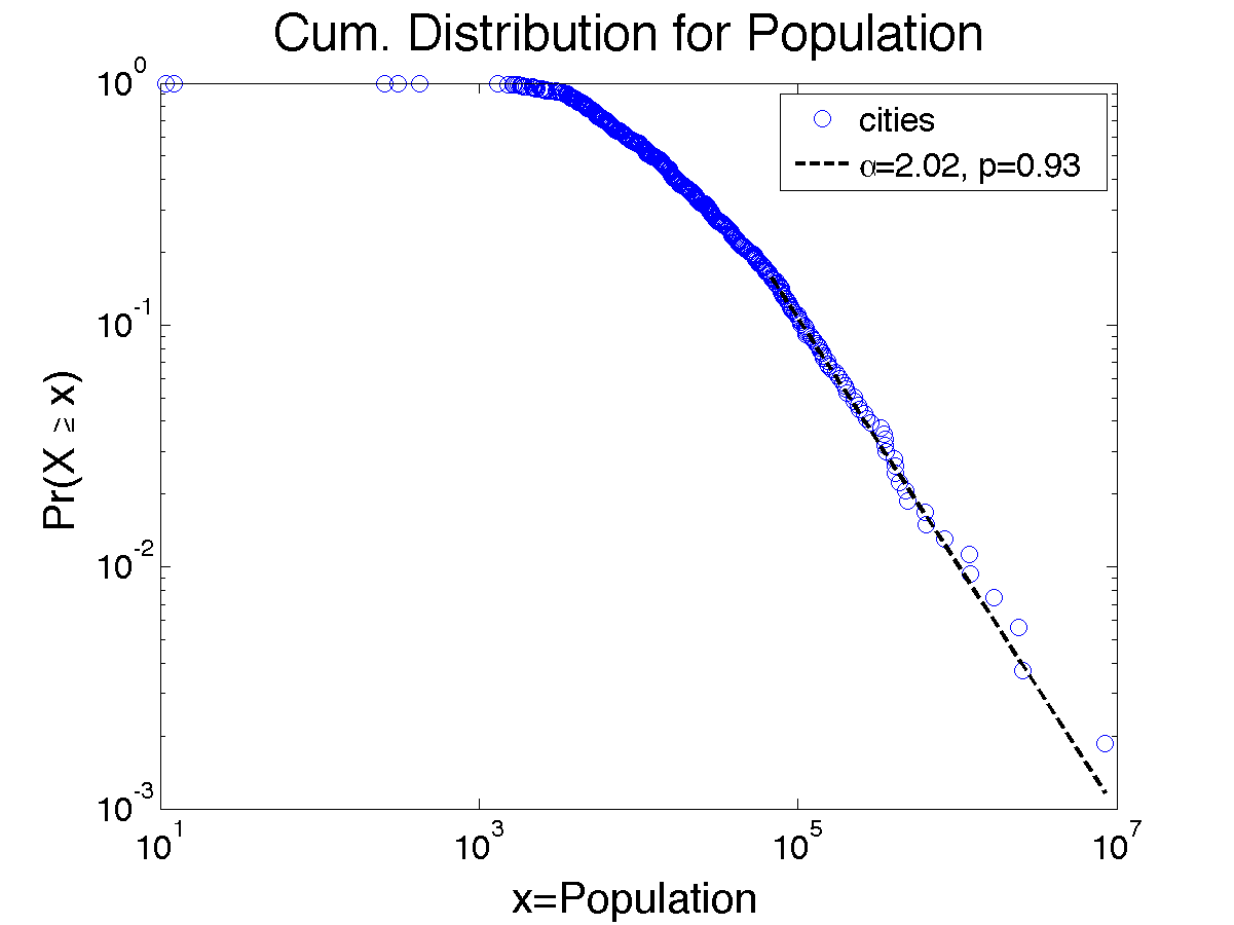}}
\caption{Zipf distribution of metropolitan areas for $\rho_c=14$prs/ha and $30\%$ commuters.}\label{Zipf_14F30}	
\end{figure}

This section gives a summary of the results presented for cities and metropolitan areas in the main text. In Fig.~3 we show that for the choice of \boldmath{$\rho_c=14$}prs/ha, the system of cities obtained is in very good agreement with the identified urbanised space observed through satellite images. The cumulative distribution for population is given by Fig.~3(b), and it clearly obeys Zipf's law. Fig.~\ref{Zipf_14F30} shows the same figure for metropolitan areas, i.e. for $\rho_c=14$prs/ha and $30\%$ commuters. We observe that Zipf's law still holds, and both distributions are very similar.
Table~\ref{table_clust14} presents the results of the analysis carried out for 30 different variables, for cities and metropolitan areas defined at $\rho_c=14$prs/ha. These results give the values and confidence intervals for the indicators presented in the main text in Fig.~5. 

\begin{table}
\centering
\begin{threeparttable} 
 \caption{Urban indicators for cities defined at a density of $14$prs/ha with and without commuters}
\begin{tabular}{lccc|ccc|c}
 Urban indicators for $\rho=14$prs/ha&\multicolumn{3}{ c| }{no commuting}& \multicolumn{3}{ c| }{30\% commuters}&\\ 
UK census 2001 &$\beta$&$95\%$ CI&$R^2$& $\beta$&$95\%$ CI&$R^2$&n. cities \\ 
\hline \\ 
Distance to work (km)&0.78&[0.76,0.81]&0.86&0.82&[0.80,0.85]&0.88&535\\
Empl. agriculture, hunting, forestry&0.86&[0.83,0.89]&	0.84&0.92&[0.89,0.95]&0.87&535\\
Area of greenspace (1000m$^2$)\tnote{*}&0.95&[0.92,0.99]&0.85&1.23&[1.19,1.28]&0.86&477\\
Area of domestic gardens (1000m$^2$)\tnote{*}&0.97&[0.95,0.99]&0.96&1&[0.99,1.02]&0.96&477\\
Area of dom. buildings (1000m$^2$)\tnote{*}&0.97&[0.96,0.98]&0.99&0.99&[0.98,1.00]&0.99&477\\
Empl. construction&0.98&[0.97,1.00]&0.97&0.99&[0.98,	1.00]&0.98&535\\
Empl. manufacturing&1&[0.97,1.02]&0.93&1&[0.97,1.02]&0.94&535\\
Area of road (1000m$^2$)\tnote{*}&1&[0.99,1.01]&0.99&1.04&[1.03,1.06]&0.98&477\\
Empl. process; plant \& machine op.&1.01&[0.98,1.03]&0.93&1&[0.98,1.03]&0.94&535\\
Empl. elementary occupations&1.01&[0.99,1.02]&0.98&1&[0.99,1.01]&0.98&535\\
Area of rail (1000m$^2$)\tnote{*} \tnote{\dag}&1.03&[0.88,1.18]&0.67&1.05&[0.90,1.20]&0.67&97\\
Area of path (1000m$^2$)\tnote{*}&1.06&[1.02,1.09]&0.88&1.06&[1.03,1.09]&0.89&477\\
\\
All household spaces&0.99&[0.98,0.99]&1&0.99&[0.98,0.99]&1&535\\
Empl. personal service occupations&0.99&[0.98,1.00]&0.99&0.99&[0.98,1.00]&0.99&535\\
All people aged 16-74 in employment&1&[0.99,1.01]&	0.99&1&[1.00,	1.01]&0.99&535\\
\\
Total number of patents 2000-2011\tnote{\dag}&0.98&[0.84,1.13]&0.67&0.99&[0.86,1.12]&0.71&100\\
Number of train stations\tnote{\dag}&0.98&[0.86,1.11]&0.73&1.04&[0.91,1.16]&0.75&100\\
Empl. hotels and restaurants&0.98&[0.97,1.00]&0.96&0.98&[0.97,1.00]&0.96&535\\
Empl. skilled trades occupations&0.99&[0.97,1.00]&0.98&0.99&[0.98,1.00]&0.98&535\\
Empl. managers and senior officials&0.99&[0.97,1.02]&0.94&1.01&[0.99,1.03]&0.95&535\\
Net income (weekly) (before housing)&0.99&[0.98,1.00]&0.98&1&[0.99,1.01]&0.99&535\\
Net income (weekly) (after housing)&0.99&[0.98,1.01]&0.98&1&[0.99,1.01]&0.99&535\\
Total income (weekly)&1&[0.99,1.02]&0.98&1.01&[0.99,1.02]&0.98&535\\
Net income (weekly)&1&[0.99,1.02]&0.98&1.01&[1.00,1.02]&0.98	&535\\
Empl. ass. prof. and technical occ.&1.01&[1.00,1.03]&0.96&1.02&[1.00,1.03]&0.96&535\\
Empl. professional occupations&1.03&[1.00,1.05]&0.9&1.03&[1.01,1.06]&0.91&535\\
Empl. real estate, business activities&1.03&[1.01,1.06]&0.93&1.03&[1.01,1.05]&0.93&535\\
Empl. sales, customer service occ.&1.03&[1.02,1.04]&0.99&1.02&[1.01,1.03]&0.99&535\\
Empl. financial intermediation&1.08&[1.04,1.11]&0.86&1.08&[1.04,1.11]&0.88&535\\
Area non-dom. buildings (1000m$^2$)\tnote{*}&1.11&[1.08,1.15]&0.88&1.13&[1.09,1.17]&0.89&477\\
\hline
\label{table_clust14}
 \end{tabular}
\begin{tablenotes}\footnotesize 
\item[*] Welsh cities are excluded, since data for Wales on infrastructure is not available
\item[$\dag$] Only considered cities with a minimum population size of $5.10^4$ individuals in order to avoid null values
\end{tablenotes}
\end{threeparttable}
 \end{table}

\section*{Sensitivity of the scaling exponent to city definitions}

In this section we extend the results of the sensitivity analysis for the scaling exponent to a dozen urban indicators. We show the results for no population size cutoff, and for a cutoff of $10^4$ individuals. The behaviour of the exponent is displayed through heatmaps covering the full parameter space. On the horizontal axis there are 40 different population density thresholds ($\rho={1,2,...,40}$prs/ha) and on the vertical axis there are 21 different percentages for commuting thresholds  ($\tau={0,5,..100}\%$ individuals).

\begin{figure}
\centerline{\includegraphics{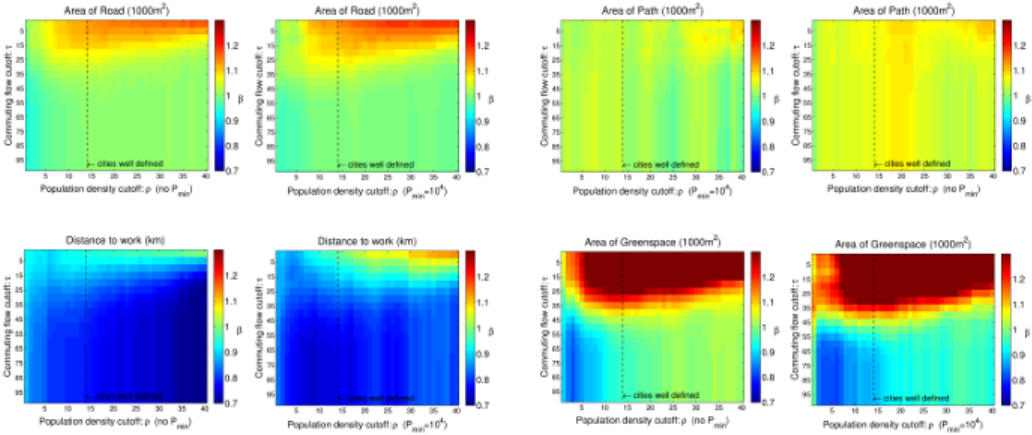}}
\caption{Heatmaps for variables whose exponent is expected to behave as $\beta<1$, such as infrastructure variables. For each variable we have 2 maps: one with no minimum population size, and another for a minimum population size of $10^4$ individuals.}	\label{sublinear_heatmaps}
\end{figure}

Let us first look at exponents of variables that should lie in the sublinear regime Fig.~\ref{sublinear_heatmaps}.
We observe nevertheless, that for two such important infrastructure variables, area of roads and area of paths, the exponent is not at all sublinear. 
From this selection, only distance to work truly belongs to the sublinear regime. 
On the other hand, area of green space clearly demonstrates the sensitivity of the exponent to city boundaries when nonlinear dependencies are present.

\begin{figure}
\centerline{\includegraphics{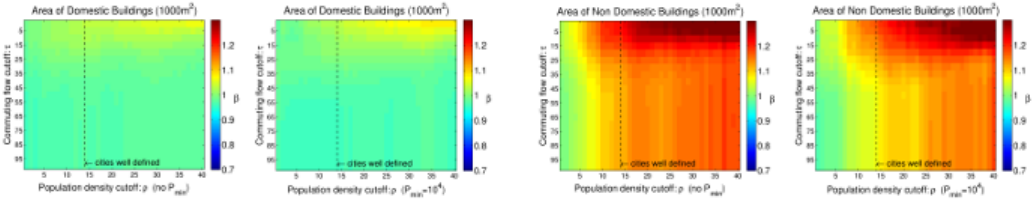}}
\caption{Heatmaps for area of domestic and non-domestic buildings.}\label{area_buildings}	
\end{figure}

Further cases of interest within the infrastructure variables are the area of domestic and of non-domestic buildings, Fig.~\ref{area_buildings}. The latter can be seen as reflecting the economic activity within cities, and once again we see how the exponent varies depending on the boundaries and distribution of cities considered.

\begin{figure}
\centerline{\includegraphics{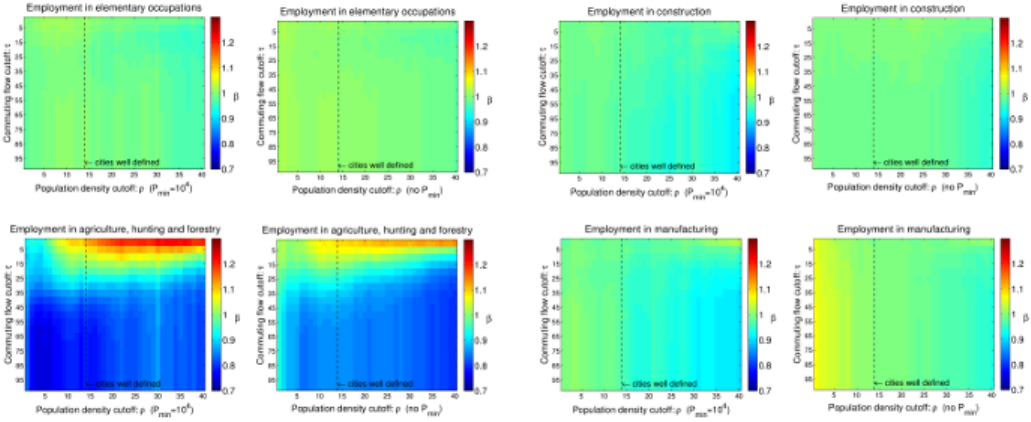}}
\caption{Heatmaps for basic employment categories.}\label{sub_employment}	
\end{figure}

\begin{figure}
\centerline{\includegraphics{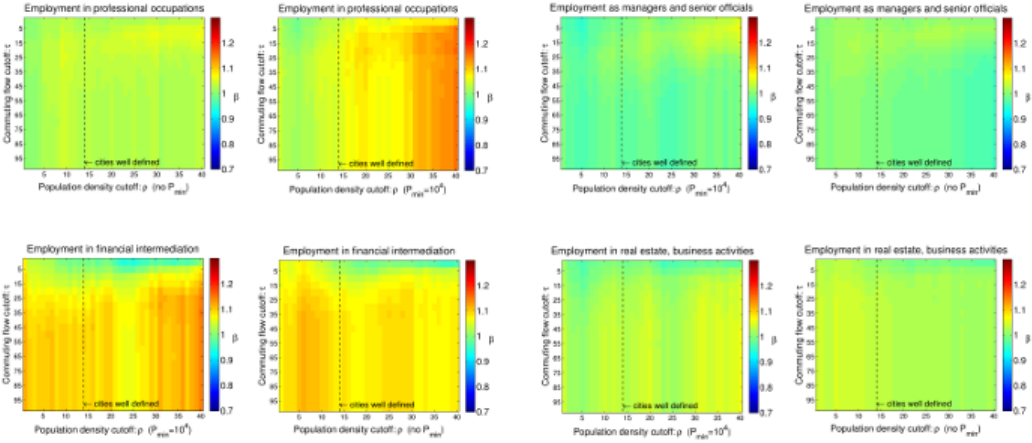}}
\caption{Heatmaps for skilled employment categories.}\label{super_employment}	
\end{figure}

Let us now turn our attention to variables corresponding to employment categories. Fig.~\ref{sub_employment} shows the heatmaps for categories expected to lie within the sublinear regime, since these correspond to employment requiring only basic skills. On the other hand, Fig.~\ref{super_employment} depicts the categories where skilled individuals are needed.

For basic employment categories, only employment in agriculture belongs to the sublinear regime. The heatmaps show that the value of the exponent is sensitive to commuting flows.
For employment in professional occupations and financial intermediation, mild superlinearity is observed.
Nevertheless, the exponent seems to belong to the linear regime for employment as managers and in real estate.
It is important to note that for employment in professional occupations, a sensitivity for city boundaries is present for a population cutoff of $10^4$ but none is apparent when no cutoff is applied.

\section*{Dragon-Kings}
We argued in the text that size alone does not provide enough information to define the state of a city according to eq.~(1). The scaling effects observed in certain countries, might pertain to a very specific  system of cities not found in E\&W.
In this particular case, London behaves as an outlier. 
This is an extremely important city that cannot be removed from the statistics of cities, as one might proceed when encountering outliers in a distribution classified as errors or biases in the data.
In this case, we are observing an outlier whose dynamics are different to the rest of the cities in the distribution.
These particular sets of large events are identified by Sornette and collaborators (see references in main text) as \emph{dragon-kings}, given that they are extremely important (kings), and different in nature to the rest of the events in the distribution (dragons). 
These might be the outcome of some sort of amplification mechanism from positive feedbacks.

\begin{figure}
\centerline{\includegraphics{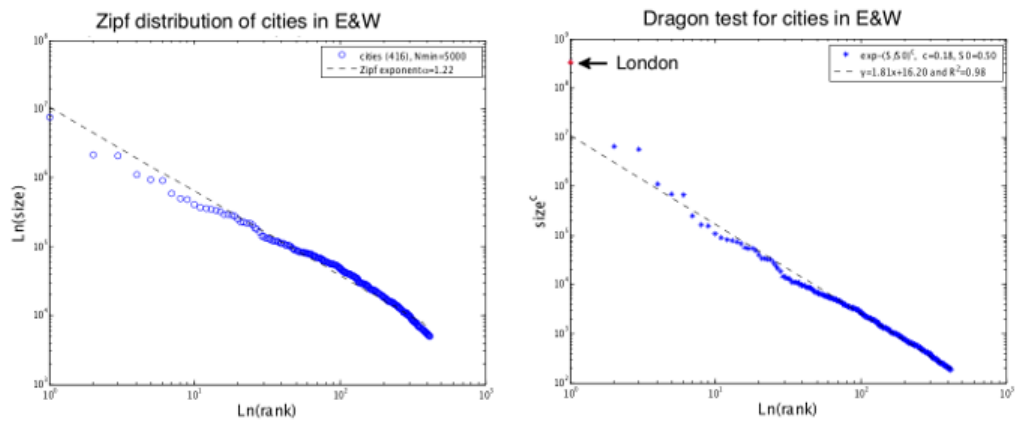}}
\caption{Cities in E\&W are defined at $\rho_c=14$prs/ha. Left: Zipf's law distribution of cities; right: transformed distribution showing London as a dragon-king.}\label{Zipf_dragon}
\end{figure}

There is no unique methodology to identify these sorts of events. 
Here we visually show that London is an outlier by employing the transformation in \cite{Sornette2009}.
Given that a power law is a special case of a stretched exponential distribution, we transform the size of cities $s$ as follows:
\beq
s \rightarrow \exp-\left(\frac{s}{s_0}\right)^c \label{s_exp}
\eeq
The usual rank size distribution of city sizes is shown in Fig.~\ref{Zipf_dragon} (left), and appears to be more or less Zipf. 
The distribution using the transformation in eq.~(\ref{s_exp}) is given in Fig.~\ref{Zipf_dragon} (right) which clearly shows London as a red point well outside the fitting line.

Further statistical tests showing that London is an outlier can be found in \cite{Pisarenko_Sornette2012}. There they employ two different tests, the DK and U-test, which consist in showing that London, as the larger event, is not generated by the same power law distribution as the rest of the cities.


\section*{Scaling results for Larger Urban Zones (E\&W)}
\begin{figure}[ht]
\centerline{\includegraphics{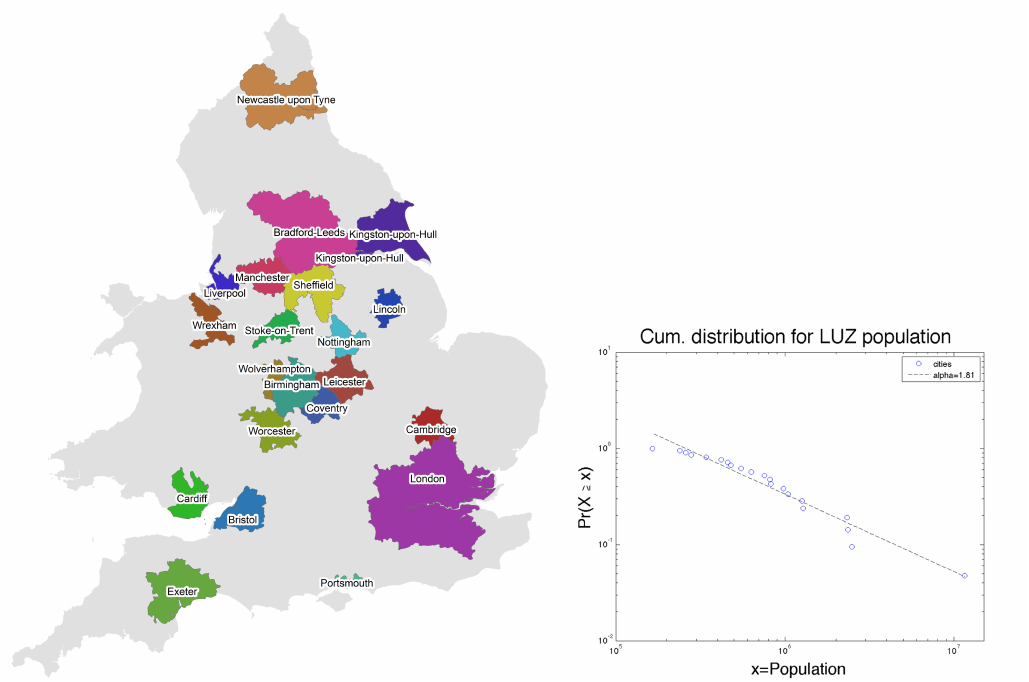}}
	\caption{Map (left) and size distribution (right) of LUZ for E\&W}\label{LUZ_map_Zipf}
\end{figure}

In order to provide a wider perspective of the behaviour of urban indicators beyond our own definition of cities, we present the results for the definition of cities in E\&W in terms of Larger Urban Zones (LUZ). LUZ were introduced by Eurostat in order to provide a consistent definition of cities across Europe. A map of the LUZ cities and a plot of their size distribution  are presented in Fig.~\ref{LUZ_map_Zipf}. 

\begin{figure}
\centering
	\includegraphics{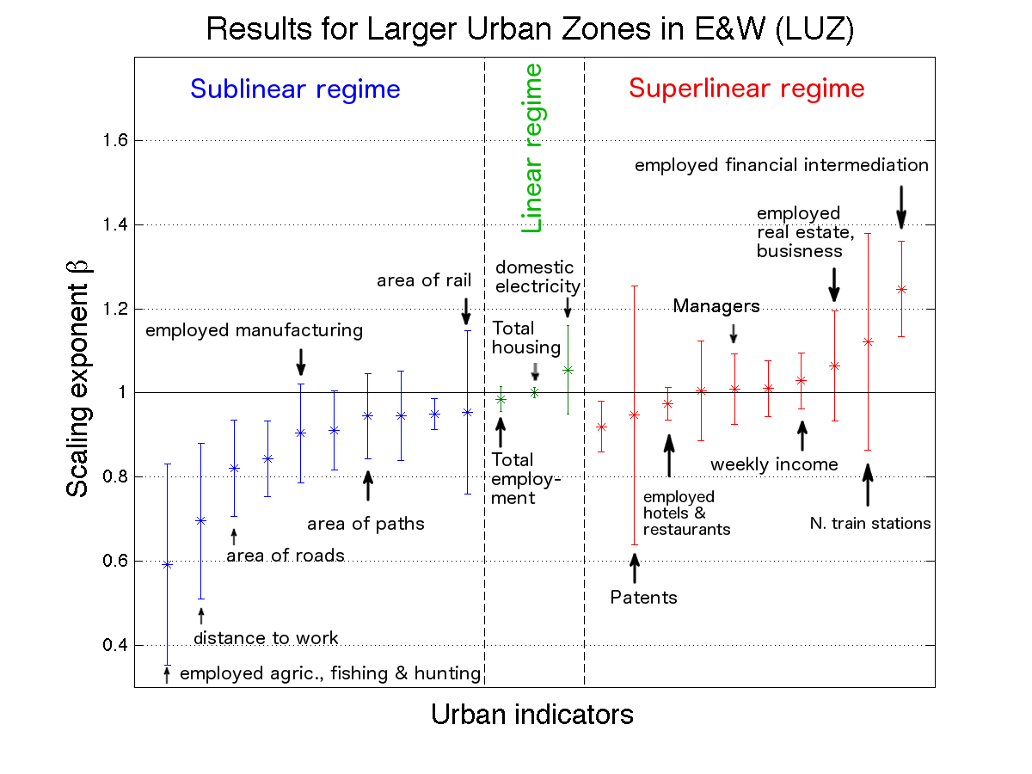}
	\caption{Exponents with 95\% CI for different urban indicators coloured-coded according to their expected regime.}\label{betas_LUZ_graph} 
\end{figure}

The analysis is undertaken by classifying each indicator in terms of the expected three regimes: sublinear, linear and superlinear. The results are shown in Fig.~\ref{betas_LUZ_graph}, and the details of the variables and their statistical properties can be found in Table~\ref{betas_LUZ}.
We observe that once again the expected scaling behaviour is not corroborated for the urban indicators lying in the superlinear regime. 

\begin{table*}
\centering
\begin{threeparttable} 
 \caption{Urban indicators for LUZ in E\&W}
\begin{tabular}{lcccc}
 Urban indicators for LUZ, UK census 2001&$\beta$&$95\%$ CI&$R^2$&n. cities \\ 
\hline \\ 
Employed in agriculture, hunting and forestry&0.59&[0.35,0.83]&0.58&21\\
Distance to work (km)&0.70&[0.51,0.88]&0.76&21\\
Area of road (1000m$^2$)\tnote{*}&0.82&[0.71,0.93]&0.93&19\\
Area of domestic gardens (1000m$^2$)\tnote{*}&0.84&[0.75,0.93]&0.96&19\\
Employed in manufacturing&0.90&[0.79,1.02]&0.93&21\\
Number of bus stops&0.91	&[0.82,1.01]&0.96&21\\
Area of path (1000m$^2$)\tnote{*}&0.94&[0.84,1.05]&0.96&19\\
Employed: process; plant and machine op. &0.95&[0.84,1.05]&0.95&21\\
Employed in elementary occupations&0.95&[0.91,0.99]&0.99&21\\
Area of rail (1000m$^2$)\tnote{*}&0.95&[0.76,1.15]&0.86&19\\
\\
All people aged 16-74 in employment&0.99&[0.96,1.02]&1.00&21\\
All household spaces&1.00&[0.99,1.01]&1.00&21\\
Consumption of domestic electricity&1.06&[0.95,1.16]&0.96&21\\
\\
Employment in skilled trades occupations&0.92&[0.86,0.98]&0.98&21\\
Total number of patents (2000-2011)&0.95&[0.64,1.26]&0.68	&21\\
Employed in hotels and restaurants&0.97&[0.94,1.01]&0.99&21\\
Employed in professional occupations&1.01	&[0.89,1.12]&0.94&21\\
Employed as managers and senior officials&1.01&[0.92,1.09]&0.97&21\\
Employment in associate prof and technical occ.&1.01&[0.94,1.08]&0.98&21\\
Total income (weekly) &1.03&[0.96,1.10]&0.98&21\\
Employed in real estate, business activities&1.06&[0.93,1.20]&0.94&21\\
Number of train stations&1.12&[0.86,1.38]&0.81&21\\
Employed in financial intermediation&1.25&[1.13,1.36]&0.96&21\\

\hline
\label{betas_LUZ}
 \end{tabular}
\begin{tablenotes}\footnotesize 
\item[*] The two Welsh cities in LUZ are excluded, since data for Wales on infrastructure is not available
\end{tablenotes}
\end{threeparttable}
 \end{table*}


\end{document}